\documentclass[10pt]{article}
\usepackage[a4paper,margin=1in]{geometry}
\usepackage{amsmath,amssymb}
\usepackage{graphicx}
\usepackage{float}
\usepackage[hidelinks]{hyperref}
\usepackage{cite}

\usepackage{booktabs}

\pdfoutput=1

\newcommand{\emit}{\textrm{emit}}
\newcommand{\obs}{\textrm{obs}}
\newcommand{\drag}{\textrm{drag}}

\newcommand{\tc}{\tilde{c}}
\newcommand{\tm}{\tilde{m}}
\newcommand{\te}{\tilde{e}}

\newcommand{\tth}{\tilde{h}}
\newcommand{\tk}{\tilde{k}}
\newcommand{\tr}{\tilde{r}}

\newcommand{\tz}{\tilde{z}}
\newcommand{\tG}{\tilde{G}}

\newcommand{\tT}{\tilde{T}}

\newcommand{\tH}{\tilde{H}}
\newcommand{\trho}{\tilde{\rho}}

\newcommand{\tomega}{\tilde{\omega}}
\newcommand{\tepsilon}{\tilde{\epsilon}}

\newcommand{\tmu}{\tilde{\mu}}
\newcommand{\tnu}{\tilde{\nu}}
\newcommand{\tlambda}{\tilde{\lambda}}

\newcommand{\GR}{\textrm{GR}}

\newcommand{\Tb}{\textrm{b}}
\newcommand{\Tm}{\textrm{m}}
\newcommand{\Tr}{\textrm{r}}

\newcommand{\Teq}{\textrm{eq}}

\title{Cosmological Time Dilation and the Hubble Tension in the meVSL Model}
\author{Seokcheon Lee\\
\small Department of Physics, Institute of Basic Science, Sungkyunkwan University, Suwon 16419, Korea\\
\small \texttt{skylee@skku.edu}\\\small ORCID: 0000-0003-0861-1300}
\date{}

\begin{document}

\maketitle

\begin{abstract}
We investigate the relation between cosmological time dilation (CTD), photon--baryon decoupling, and the Hubble tension in the minimally extended varying--speed--of--light (meVSL) framework. The generalized cosmological time (GCT) framework distinguishes cross-epoch temporal standardization from local interaction laws. For the diagnostic early-Universe calculation, the hydrogen recombination history is taken to be GR-like, while the homogeneous meVSL opacity prescription retains the correlated $b$ dependence of the dimensional Thomson cross section. This background scaling of the dimensional Thomson cross section does not imply that the locally measured dimensionless fine-structure constant $\alpha$ varies with cosmological epoch. Keeping the rapid variation of the free-electron fraction $X_e(z)$ in the decoupling condition gives the local representation $z_\ast\simeq1090-315b$ over the range relevant here. The baryon drag optical depth is determined by the Thomson momentum-transfer rate $\dot\kappa/R$, with $R=3\rho_b/(4\rho_\gamma)$, and accumulates from the present epoch to redshift $z$. Positive $b$ then shifts the drag epoch slightly downward in redshift. For example, $b=0.03$ gives $z_{\rm drag}\simeq1052.2$ and $r_{\rm drag}\simeq147.84~\mathrm{Mpc}$, compared with $r_{\rm drag}\simeq147.15~\mathrm{Mpc}$ at $b=0$ in the same calculation. The drag-epoch sound horizon remains nearly unchanged and does not yield the several-percent reduction required for a sound-horizon explanation of the Hubble tension. In the time-domain sector, same-event proper-time dilation remains $1+z$, whereas standardizing intrinsic durations across epochs in GCT gives the conditional relation $n=1-b/4$. Consequently, CTD and the acoustic ruler provide distinct empirical tests of the temporal framework.
\end{abstract}

\newpage 

\tableofcontents 

%=======================================================
\section{Introduction} %sec:1

The standard model of cosmology (SMC) is based on general relativity (GR) and the Friedmann--Lema\^itre--Robertson--Walker (FLRW) metric. The FLRW metric follows from the cosmological principle (CP)---spatial homogeneity and isotropy---together with Weyl's postulate, which defines a global cosmic time~\cite{Ryder09,Islam01,Narlikar02,Hobson06,Roos15}. With the inclusion of a cosmological constant $\Lambda$ and cold dark matter (CDM), this framework explains a wide range of observations, including the cosmic microwave background (CMB), large-scale structure (LSS), and Type Ia supernovae (SNe Ia).

However, persistent discrepancies remain between early- and late-time determinations of key parameters, most prominently the Hubble constant $H_0$~\cite{Perivolaropoulos:2021jda,Abdalla:2022yfr,DiValentino:2025sru}. These ``cosmological tensions'' have been attributed to possible systematics~\cite{Efstathiou:2020wxn,Schosser:2024vbj,Perivolaropoulos:2024yxv}, statistical fluctuations~\cite{Bernal:2016gxb,Raveri:2019mxg}, or new physics beyond $\Lambda$CDM~\cite{Poulin:2018cxd,Jedamzik:2020krr}. In this work, we explore how the minimally extended varying-speed-of-light (meVSL) model~\cite{Lee:2020zts,Lee:2022heb,Lee:2023rqv,Lee:2023ucu,Lee:2024nya,Lee:2024kxa,Lee:2023bjz,Lee:2024mal,Lee:2024zcu,Lee:2025osx} affects both early-Universe standard-ruler physics and cosmological time-domain observables.

A particularly direct probe of cosmic time is provided by cosmological time dilation (CTD) in SNe Ia~\cite{Leibundgut:1996qm,SupernovaSearchTeam:1997gem,Foley:2005qu,Blondin:2007ua,Blondin:2008mz,DES:2024vgg}. Empirically, the observed duration of a supernova light curve scales as
\begin{equation}
    \Delta t_{\rm obs} = (1+z)^n \, \Delta t_{\rm emit} \, ,
    \label{CTD}
\end{equation}
with $n=1$ for the standard proper-time relation. In the generalized cosmological time (GCT) framework adopted here, a departure of the fitted CTD exponent from unity can be related to the meVSL parameter $b$ only after intrinsic durations at different epochs are expressed in a common GCT temporal standard. For the homogeneous temporal scaling used in this work, the conditional mapping is $n=1-b/4$~\cite{Lee:2020zts,Lee:2024mal,Lee:2024zcu,Lee:2025osx}.

The same parameter $b$ also enters the pre-recombination background and the effective Thomson-opacity prescription adopted below, so photon decoupling, baryon drag, and the sound horizon provide a complementary consistency test. We examine quantitatively how these early-Universe quantities respond to $b$ and compare that response with the independent time-domain test supplied by CTD.

In this manuscript, we first review the modified Friedmann equations in the meVSL framework (Section~\ref{sec:FE}) and quantify the resulting photon-decoupling, baryon-drag, and sound-horizon behavior (Section~\ref{sec:tension}). We then formulate the supernova time-dilation observable within the GCT framework and summarize current Dark Energy Survey (DES) constraints (Section~\ref{sec:CTD}). Finally, Section~\ref{sec:conclusion} summarizes the implications for the relation between time-domain observables and the Hubble tension.

\section{Modified Friedmann Equations (Summary)}
\label{sec:FE} %sec:2

In the meVSL model, the Einstein equations for a homogeneous and isotropic universe lead to the following modified Friedmann equations, as derived in our earlier work~\cite{Lee:2020zts,Lee:2024mal,Lee:2024zcu,Lee:2025osx}
\begin{align}
&\tH^2 \equiv H^2 a^{\frac{b}{2}} = \frac{8\pi \tG}{3} \sum_i \trho_i +  \frac{\Lambda \tc^2}{3} -  \frac{k \tc^2}{a^2} \,, \label{tG00mp_summary} \\
&\frac{\ddot{a}}{a} = -\frac{4\pi \tG}{3} \sum_i (1 + 3\tomega_i)\,\trho_i + \frac{\Lambda \tc^2}{3} + \tH^2 \frac{d \ln \tc}{d \ln a} \,, \label{t3G11mG00mp_summary}
\end{align}
where the Bianchi identity gives
\begin{align}
\trho_i \tc^2 = \rho_{i0} c_0^2 a^{-3(1 + \tomega_i)} \,, \label{rhomp}
\end{align}
with $\rho_{i0}$ denoting the present-day value of the mass density of the $i$-th component. 
Here and throughout this section, quantities denoted with a tilde (e.g.\ $\tc$, $\tG$, $\trho_i$) correspond to their values in the meVSL framework, which generally scale with the cosmic scale factor $a(t)$ as shown in Table~\ref{tab:constants}. In the special case $b=0$, all tilded quantities reduce to their standard GR counterparts, such that $\tc \to c_0$, $\tG \to G_0$, $\trho_i \to \rho_i$. Untilded symbols are used to denote the conventional quantities of the $\Lambda$CDM model. 
With this notation, the equations reduce to the standard Friedmann equations when $\tc = c_0$. 
The additional derivative term proportional to $d \ln \tc / d \ln a$ captures the effect of a time-dependent effective speed of light on the cosmic acceleration.

Expressed in terms of present-day parameters, the first Friedmann equation can be written as
\begin{align}
\tH^2 &= \left[ \frac{8\pi \tG_0}{3} \sum_i \rho_{i0}\,a^{-3(1+\omega_i)} +  \frac{\Lambda c_0^2}{3} - k\,\frac{c_0^2}{a^2} \right] \frac{\tc^2}{c_0^2} 
\equiv H^2 \frac{\tc^2}{c_0^2} \,, \label{H2me_summary}
\end{align}
where $H$ denotes the standard Hubble parameter in GR. The corresponding acceleration equation becomes
\begin{align}
\frac{\ddot{a}}{a} &= \left( -\frac{4 \pi G_0}{3} \sum_i (1 + 3 \omega_i) \rho_{i0} a^{-3(1+\omega_i)} + \frac{\Lambda c_0^2}{3} + H^2 \frac{d \ln \tc}{d \ln a} \right) \frac{\tc^2}{c_0^2} \,, \label{ddotaoa_summary}
\end{align}
making explicit how both the expansion rate and acceleration are modified through $\tc(a)$.

As a direct consequence, a useful relation follows by isolating $\tc/\tH$ from Eq.~\eqref{H2me_summary}:
\begin{align}
\frac{\tc}{\tH} = \frac{c_0}{H} \,, \label{H2me2_summary}
\end{align}
which shows that the ratio $\tc/\tH$, and hence the corresponding comoving Hubble radius at fixed $a$, retains its standard value under the meVSL scaling~\cite{Petit:1988ti,Petit:1988-2,Petit:1989,Midy:1989,Barrow:1998eh,Moffat:1992ud,Petit:1995ys,Albrecht:1998ir,Barrow:1998he,Clayton:1998hv,Barrow:1999jq,Clayton:1999zs,Brandenberger:1999bi,Bassett:2000wj,Gopakumar:2000kp,Magueijo:2000zt,Magueijo:2000au,Magueijo:2003gj,Magueijo:2007gf,Petit:2008eb,Roshan:2009yb,Sanejouand:2009zz,Nassif:2012dr,Moffat:2014poa,Ravanpak:2017kdg,Costa:2017abc,Cruz:2018avc}. Inflation or another mechanism is therefore still required to explain early-universe causal contact.

Beyond the background dynamics, the meVSL model consistently embeds local physics within its scaling relations~\cite{Lee:2020zts,Lee:2025osx}. Quantities arising in special relativity, electromagnetism, and thermodynamics exhibit apparent cosmological evolution without requiring any change in the underlying local laws of physics. Table~\ref{tab:constants} summarizes representative scaling relations of physical quantities and constants, showing the correlated cosmological scalings used to retain the corresponding local relations.

\begin{table}[htbp]
\centering
\resizebox{\textwidth}{!}{%
\begin{tabular}{l l l l}
\toprule
\textbf{Local Physical Laws} & \textbf{Special Relativity} & \textbf{Electromagnetism} & \textbf{Thermodynamics} \\
\midrule
\textbf{Quantities} & 
$\tm = m_0 a^{-b/2}$ & 
$\te = e_0 a^{-b/4}, \ \tlambda = \lambda_0 a, \ \tnu = \nu_0 a^{-1+b/4}$ & 
$\tT = T_0 a^{-1}$ \\
\textbf{Constants} & 
$\tc = c_0 a^{b/4}, \ \tG = G_0 a^b$ & 
$\tepsilon = \epsilon_0 a^{-b/4}, \ \tmu = \mu_0 a^{-b/4}$ & 
$\tk_{\text{B}} = k_{\text{B}0}, \ \tth = h_0 a^{-b/4}$ \\
\textbf{Energies} & 
$\tm \tc^2 = m_0 c_0^2$ & 
$\tth \tnu = h_0 \nu_0 a^{-1}$ & 
$\tk_{\text{B}} \tT = k_{\text{B}} T_0 a^{-1}$ \\
\bottomrule
\end{tabular}%
}
\caption{Apparent cosmological scaling of physical quantities and constants in the meVSL model. Subscript $0$ denotes present-day measured values; tilded quantities correspond to meVSL scalings, reducing to their GR values when $b=0$.}
\label{tab:constants}
\end{table}

\section{Observational Consequences in the meVSL model}
\label{sec:tension} %sec:3

In the meVSL model, cosmological observables acquire a mild dependence on the parameter $b$ that characterizes the effective scaling of the speed of light. Crucially, these dependencies do not arise from new fundamental interactions, but from a modified effective description of cosmic time~\cite{Lee:2020zts,Lee:2024mal,Lee:2024zcu,Lee:2025osx}. As a result, the photon-decoupling and baryon-drag epochs, the sound-horizon scale, and the fitted CTD exponent can acquire $b$-dependent shifts in the cosmological representation.

A wide range of probes are sensitive to such effects, including Big Bang Nucleosynthesis (BBN), the CMB anisotropies, baryon acoustic oscillations (BAO), SNe Ia, direct measurements of $\tH(\tz)$, and the propagation of gravitational waves (GWs). Observational limits on the time variation of the fine-structure constant $\alpha$ constrain the locally realized dimensionless electromagnetic coupling. In the GCT interpretation, such local bounds must be distinguished from correlated cross-epoch scalings of dimensional cosmological quantities and from the homogeneous temporal normalization~\cite{Lee:2023xfg}. The implications of meVSL for these and other observables have been discussed in previous work~\cite{Lee:2020zts,Lee:2023rqv,Lee:2023ucu,Lee:2024nya,Lee:2024kxa}. Here we focus on whether the same background parameter that enters the CTD standardization prescription also produces a significant change in the early-Universe acoustic ruler.

\subsection{Connection to the \texorpdfstring{$H_0$}{H0} tension}

It is well known that attempts to address the discrepancy between CMB-inferred and local distance-ladder determinations of $H_0$ often motivate modifications to early-Universe physics, particularly at or before recombination. Models that effectively reduce $r_{\rm drag}$ provide one possible pathway to bringing these determinations into closer agreement~\cite{Evslin:2017qdn,Chiang:2018xpn,Poulin:2018cxd,Aylor:2018drw,Agrawal:2019lmo,Knox:2019rjx,Sekiguchi:2020teg,Lee:2022gzh,Schoneberg:2024ynd,Chatrchyan:2024xjj,Mirpoorian:2024fka,Smith:2025zsg}. The meVSL model modifies the pre-recombination expansion and Thomson-scattering sector through the same parameter $b$ that appears in its temporal scaling~\cite{Lee:2020zts,Lee:2025aha}. We examine explicitly whether the resulting shifts of photon decoupling and baryon drag are large enough to reduce the acoustic ruler by the amount required for such an interpretation.

\subsubsection{Sound horizon from BAO}

The comoving size of the sound horizon at the drag epoch ($\tz_{\drag}$), given by~\cite{Lee:2020zts,Lee:2025aha}
\begin{align}
\tr_{\drag} 
\equiv \int_{0}^{t_{\drag}}\frac{ \tc_{s}(t) }{a(t)} \, dt
= \int_{0}^{a_{\drag}} \frac{\tc_s(a)}{a^2 \tH(a)} \, da 
= \int_{z_{\drag}}^{\infty} \frac{\tc_{s}(z)}{\tH(z)} \, d z \,,  
\label{rsd} 
\end{align}
represents the maximum distance that an acoustic wave could have propagated in the primordial photon-baryon plasma from the Big Bang up to the time when baryons were released from the Compton drag of photons — \textit{i.e.}, the end of the baryon drag epoch. This characteristic scale is imprinted on the matter power spectrum and serves as a standard ruler for cosmological observations.
For definiteness, the early-Universe calculation considered here takes the hydrogen recombination history to be GR-like, $X_e(z,b)=X_e^{\GR}(z)$, while the effective Thomson cross section entering the homogeneous cosmological opacity scales as in Eq.~\eqref{tsigmaT}. Here $\tilde\sigma_{\rm T}$ is the dimensional cross section expressed in the common meVSL/GCT background standards used for the cosmological opacity calculation. Its correlated $b$ dependence is not identified with a variation of the locally measured dimensionless fine-structure constant $\alpha$; on each local hypersurface the interaction law retains its standard form. This prescription isolates the stated meVSL opacity scaling without assigning an additional $b$ dependence to the full atomic recombination network. A precision treatment in which the local atomic rates themselves are modified would require a separately specified, self-consistent $b$-dependent recombination code.

For the baryon--photon plasma, we use the meVSL sound-speed prescription
\begin{align}
\tc_s^2 
&\equiv \frac{\partial \tilde P_{\gamma}}{\partial \tilde\rho_{\gamma + \Tb}}  
= \frac{c_{0}^2}{3} (1+z)^{-\frac{b}{2}} 
\left( 1 + \frac{(3 + b/2) \trho_{\Tb}}{(4 + b/2) \trho_{\gamma}} \right)^{-1} \equiv c_{s}^{2} (1+z)^{-\frac{b}{2}} 
 \frac{1 + R}{1 + \frac{1 + b/6}{1 + b/8} R} \,, \nonumber \\
&\text{where} \quad R = \frac{3 \rho_{\Tb 0}}{4 \rho_{\gamma 0}} \frac{1}{1+z} \,.
\label{cs2}
\end{align}
The factor $(1+z)^{-b/2}$ arises from the scaling of $\tc^2$ in the meVSL model, which directly impacts the effective sound speed. Combining Eqs.~\eqref{rsd} and \eqref{cs2}, we obtain
\begin{align}
\tr_{\drag}
&= \frac{c_0}{\sqrt{3} H_0} \int_{z_{\drag}}^{\infty} 
\frac{dz}{\sqrt{\Omega_{\Tr 0} (1+z)^4 + \Omega_{\Tm 0} (1+z)^3 + \Omega_{\Lambda}}} 
\left( 1 + \frac{(3 + b/2) \trho_{\Tb}}{(4 + b/2) \trho_{\gamma}} \right)^{-1/2} \nonumber \\
&\equiv \frac{2997.92}{\sqrt{3}} 
\int_{z_{\drag}}^{\infty} \frac{d z} {Eh(z)} 
\left( 1 + \frac{(3 + b/2) \trho_{\Tb}}{(4 + b/2) \trho_{\gamma}} \right)^{-1/2} \, \mathrm{Mpc} \,,
\label{rsd2}
\end{align}
where $Eh(z) \equiv E(z) h = \sqrt{ \Omega_{\Tr 0} h^2 (1+z)^4 + \Omega_{\Tm 0} h^2 (1+z)^3 + \Omega_{\Lambda} h^2 }$. We used Eq.~\eqref{H2me_summary} and the relation $H_0 = 100\, h$ km/s/Mpc. The numerical factor $2997.92$ is $c_0/(100\,\mathrm{km\,s^{-1}\,Mpc^{-1}})$ expressed in Mpc; the factor $1/\sqrt{3}$ is displayed separately in Eq.~\eqref{rsd2}. The cosmological parameters are taken from Planck 2018 TT, TE, EE + lowE + lensing 68 \% limits~\cite{Planck:2018vyg}, with 
\[
\Omega_{\Tm 0} h^2 = 0.1423 \pm 0.0017, \quad 
\Omega_{\Tb 0} h^2 = 0.02237 \pm 0.00015, \quad 
z_{\drag} = 1059.94 \pm 0.30 \,.
\]
For reference, Planck 2018 quotes
\[
r_{\drag} = 147.09 \pm 0.26~\mathrm{Mpc}, \quad h = 0.6736 \pm 0.0054 \,.
\]
With the same fiducial parameters, the simplified integral used here gives $r_{\drag}(0)=147.154~\mathrm{Mpc}$, close to the Planck reference value.

\subsubsection{Decoupling redshift in meVSL}

Photon decoupling occurs when the Thomson scattering rate falls below the Hubble expansion rate. The opacity prescription is that specified above~\cite{Lee:2020zts}. One has
\begin{align}
\tilde\sigma_{\rm T}
= \frac{8\pi}{3}\!\left(\frac{\tilde e^2}{4\pi\tilde\varepsilon\,\tilde m_e\,\tilde c^2}\right)^{\!2}
= \sigma_{\rm T}\,(1+z)^{\frac{b}{2}} \label{tsigmaT} \,,
%\qquad \tilde c = c_0\,a^{\frac{b}{4}},
\end{align}
so that the per–photon scattering rate scales as
\begin{align}
\tilde\Gamma_{\rm T} = \tilde n_e\,\tilde\sigma_{\rm T}\,\tilde c
= \Gamma_{\rm T}\,(1+z)^{\frac{b}{4}} \label{tGamma},
\end{align}
where $\Gamma_{\rm T}$ denotes the $\Lambda$CDM rate. 
Decoupling is defined by $\tilde\Gamma_{\rm T}(\tz_\ast)=\tilde H(\tz_\ast)$. 
Using $\tilde H =H(1+z)^{-b/4}$, this condition can be written as
\begin{align}
\tilde\Gamma_{\rm T}=\tilde H
\quad\Longleftrightarrow\quad
\Gamma_{\rm T} = H\, (1+z)^{-\frac{b}{2}}.
\end{align}
Near decoupling the expansion is governed by matter and radiation, hence
\begin{align}
H(z) \simeq H_0\sqrt{\Omega_{m0}(1+z)^{3}+\Omega_{r0}(1+z)^{4}}
= H_0\sqrt{\Omega_{m0}}\,(1+z)^{\frac{3}{2}}
\left(1+\frac{1+z}{1+z_{\rm eq}}\right)^{\!\frac{1}{2}},
\end{align}
and therefore
\begin{align}
\tilde H(z_\ast)=H(z_\ast)(1+z_\ast)^{-\frac{b}{4}}
\simeq
H_0\sqrt{\Omega_{m0}}\,(1+z_\ast)^{\frac{3}{2}-\frac{b}{4}}
\left(1+\frac{1+z_\ast}{1+z_{\rm eq}}\right)^{\!\frac{1}{2}}.
\label{eq:Hz_mevsl}
\end{align}
The electron number density remains as GR~\cite{Lee:2020zts,Lee:2022heb}  
\begin{align}
n_e(z_\ast)=X_e(z_\ast)\,n_b(z_\ast)=X_e(z_\ast)\,n_{b0}(1+z_\ast)^3,
\end{align}
with $X_e$ the free electron fraction. Combining the above, the decoupling condition yields
\begin{align}
\frac{\tilde\Gamma_{\rm T}(z_\ast)}{\tilde H(z_\ast)}
=\frac{3\,\sigma_{\rm T}\,c_0\,H_0}{8\pi G_0 m_p}\,
X_e(z_\ast)\,
\frac{\Omega_{b0}h^2}{\sqrt{\Omega_{m0}h^2}}\,
(1+z_\ast)^{\frac{3+b}{2}}
\!\left(1+\frac{1+z_\ast}{1+z_{\rm eq}}\right)^{-\frac{1}{2}}
=1,
\label{eq:zstar_condition}
\end{align}
which, within the adopted opacity prescription, shifts the solution toward a lower decoupling redshift for $b>0$. The free-electron fraction changes rapidly across recombination, however, and must not be treated as a fixed prefactor when the solution for $z_\ast$ is displaced. Writing $y\equiv1+z_\ast$ and differentiating the logarithm of Eq.~\eqref{eq:zstar_condition} gives
\begin{align}
\left.\frac{dy}{db}\right|_{b=0}
= -\frac{\tfrac12\ln y_0}
{\left.\dfrac{d\ln X_e}{dy}\right|_{y_0}
+\dfrac{3}{2y_0}
-\dfrac{1}{2(1+z_{\Teq}+y_0)}}\,.
\label{eq:zstar_implicit_derivative}
\end{align}
The derivative of $X_e$ dominates the denominator near $z\simeq1090$. Using the GR recombination history described in Appendix~\ref{app:Xe}, the numerical roots of Eq.~\eqref{eq:zstar_condition} are well represented locally by
\begin{align}
z_{\ast}(b) \simeq 1090 - 315\,b,
\qquad 0\le b\lesssim0.03,
\label{tzb}
\end{align}
with the numerical solution preferred outside this local range. Representative roots are $z_\ast\simeq1086.85$, $1083.73$, and $1080.64$ for $b=0.01$, $0.02$, and $0.03$, respectively. The left panel of Figure~\ref{fig-1} shows the numerical roots together with the local fit. The weak $b$ dependence follows from the steep variation of $X_e(z)$ across the recombination transition.

\subsubsection{Drag epoch in meVSL}

The baryon drag epoch is controlled by the Thomson \emph{momentum-transfer} rate in the baryon Euler equation. With the conventional inertia ratio
\begin{equation}
R(z)\equiv\frac{3\rho_b(z)}{4\rho_\gamma(z)},
\end{equation}
the Thomson collision term is proportional to $\dot\kappa(\theta_\gamma-\theta_b)/R$~\cite{Ma:1995ey,Hu:1995en}. Correspondingly, the standard drag optical depth uses $1/R$, not $1/(1+R)$~\cite{Planck:2013pxb}. It is convenient to define a positive redshift opacity
\begin{equation}
\frac{d\tilde\tau_+}{dz}
\equiv
\frac{\tilde n_e(z)\,\tilde\sigma_{\rm T}(z)\,\tilde c(z)}
{(1+z)\,\tilde H(z)}>0.
\end{equation}
The drag epoch is then determined by
\begin{align}
\tilde\tau_{\drag}(z)
&=\int_0^z dz'\,\frac{1}{R(z')}\,
\frac{d\tilde\tau_+}{dz'}=1.
\label{eq:taudrag_def_mevsl_final}
\end{align}
Using $\tilde n_e=X_e n_{b0}(1+z)^3$, $\tilde\sigma_{\rm T}=\sigma_{\rm T}(1+z)^{b/2}$, and the exact ratio $\tilde c/\tilde H=c_0/H$, the positive differential opacity becomes
\begin{align}
\frac{d\tilde\tau_+}{dz}
=\frac{X_e(z)\,n_{b0}\,\sigma_{\rm T}\,c_0}{H(z)}
(1+z)^{2+\frac b2}.
\label{eq:dtaudz_mevsl_final}
\end{align}
At fixed $z$, positive $b$ increases the accumulated drag optical depth. Because the physical accumulation in Eq.~\eqref{eq:taudrag_def_mevsl_final} runs from $0$ to $z$, the unity condition is therefore reached at a \emph{lower} redshift. Solving Eq.~\eqref{eq:taudrag_def_mevsl_final} with the GR-like $X_e(z)$ history gives
\begin{align}
(b,z_{\drag}) = \{&(0,1059.94),(0.005,1058.65),(0.010,1057.36),
(0.015,1056.07),\\
&(0.020,1054.79),(0.030,1052.23),(0.040,1049.68)\}.
\label{eq:zdrag_nodes}
\end{align}
For interpolation of the numerical nodes over $0\le b\le0.08$, a quadratic representation is
\begin{equation}
z_{\drag}^{\rm (fit)}(b)
=1059.939-258.365\,b+48.720\,b^2,
\label{eq:zdrag_fit_exp_final}
\end{equation}
where the numerical precision of the fit should not be confused with the physical accuracy of the approximate recombination treatment.

\begin{figure}[ht]
\centering
\includegraphics[width=0.48\textwidth]{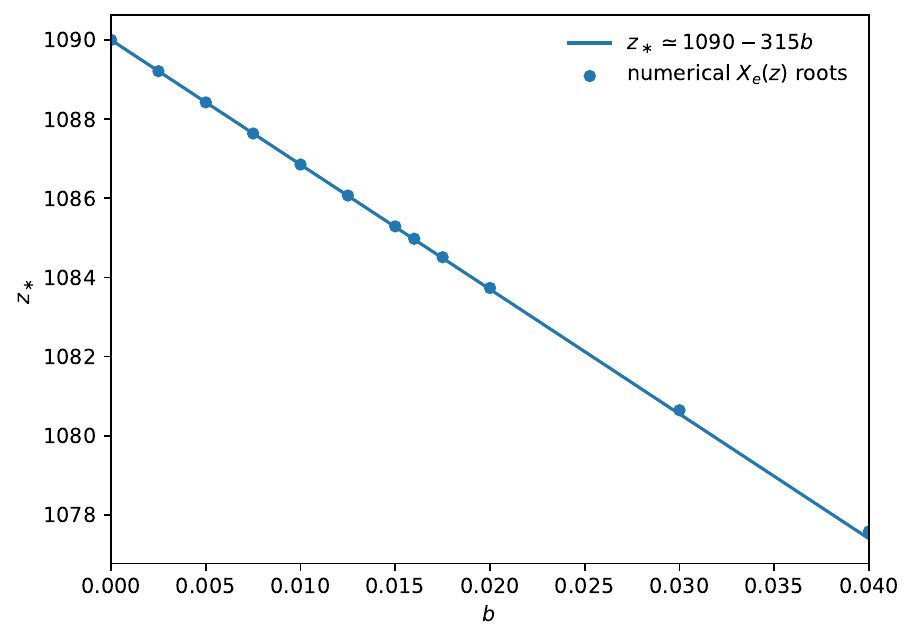}
\hfill
\includegraphics[width=0.48\textwidth]{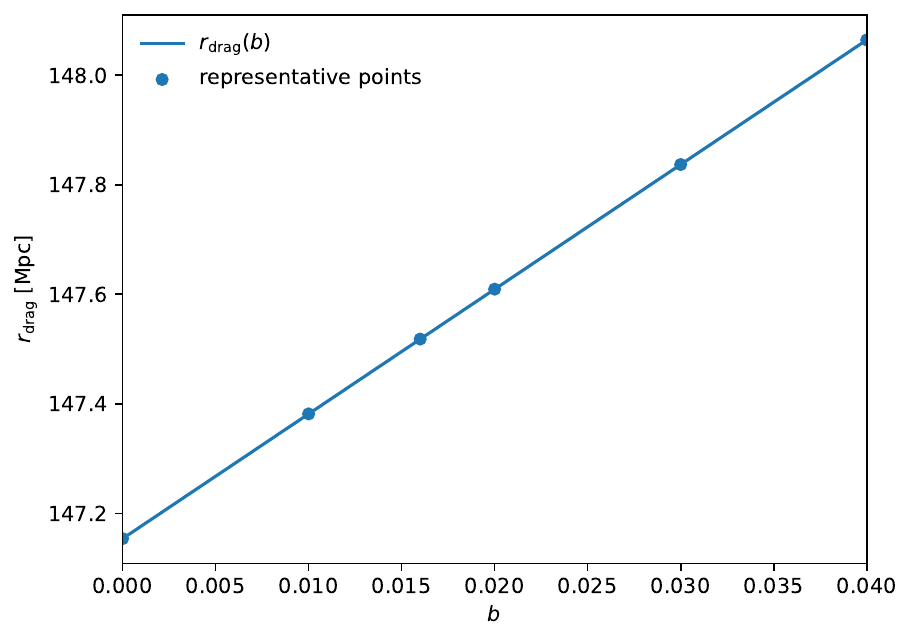}
\vspace{0.5em}
\caption{\textbf{Left:} Photon decoupling redshift from the numerical $X_e(z)$ roots of Eq.~\eqref{eq:zstar_condition}, together with the local representation $z_\ast\simeq1090-315b$. \textbf{Right:} Drag-epoch sound horizon obtained from the baryon momentum-transfer weighting $1/R$ and the corresponding $z_{\drag}(b)$. The resulting $r_{\drag}$ is nearly unchanged and increases slightly over the displayed range.}
\label{fig-1}
\end{figure}

Inserting $z_{\drag}(b)$ into Eq.~\eqref{rsd2} gives only a weak change in the acoustic ruler. Figure~\ref{fig-1} summarizes the decoupling and sound-horizon behavior.

\subsubsection{Implications for the Hubble tension}

BAO provide a standard ruler that depends on $r_{\drag}$. Anisotropic BAO analyses constrain $r_{\drag}/D_M(z)$ in the transverse direction and $r_{\drag}H(z)/c_0$ in the line-of-sight direction~\cite{Hogg:1999ad,Eisenstein:2006nk,Blake:2011en,BOSS:2013rlg}. The comoving transverse separation of a galaxy pair at redshift $z$ with angular separation $\theta$ is $D_M(z)\theta$, where the comoving angular-diameter distance is
\begin{align}
D_M(z)=\int_0^z\frac{\tilde c(z')}{\tilde H(z')}dz'
=\frac{c_0}{H_0}\int_0^z\frac{dz'}{E(z')},
\label{DM}
\end{align}
as in the standard distance-redshift relation~\cite{Hogg:1999ad}. Thus, the homogeneous lapse factor cancels between $\tilde c$ and $\tilde H$ in this distance integral. Planck and BAO analyses are sensitive to the acoustic ruler through the usual geometric combinations, while any substantial upward shift of an inverse-ruler estimate of $H_0$ would require a several-percent reduction of $r_{\drag}$. The calculation below does not produce such a reduction.

Representative values from the same simplified sound-horizon integral are listed in Table~\ref{tab:drag_results}.
\begin{table}[H]
\centering
\begin{tabular}{ccccc}
\toprule
$b$ & $z_\ast^{(\textrm{fit})}$ & $z_{\drag}$ & $r_{\drag}$ [Mpc] & $H_0^{\rm inv}$ [km s$^{-1}$ Mpc$^{-1}$]\\
\midrule
0     & 1090.00 & 1059.94 & 147.154 & 67.400\\
0.010 & 1086.85 & 1057.36 & 147.382 & 67.296\\
0.016 & 1084.96 & 1055.82 & 147.518 & 67.233\\
0.020 & 1083.70 & 1054.79 & 147.609 & 67.192\\
0.030 & 1080.55 & 1052.23 & 147.837 & 67.089\\
0.040 & 1077.40 & 1049.68 & 148.064 & 66.985\\
\bottomrule
\end{tabular}
\caption{Representative early-Universe quantities for the meVSL prescription considered here. The $z_\ast^{(\textrm{fit})}$ column uses the local representation $z_\ast\simeq1090-315b$; the $b=0.04$ entry is a short extrapolation beyond the stated local-fit range $b\lesssim0.03$. The last column is an illustrative inverse-ruler scaling, $H_0^{\rm inv}=67.4\,r_{\drag}(0)/r_{\drag}(b)$, not a replacement for a full CMB likelihood.}
\label{tab:drag_results}
\end{table}
As summarized in Table~\ref{tab:drag_results}, the sound horizon changes from $147.15$ Mpc at $b=0$ to about $147.84$ Mpc at $b=0.03$, corresponding to only a $\simeq0.46\%$ increase. Even at $b=0.04$ the change is only $\simeq0.62\%$. The acoustic ruler changes only weakly over the range of interest, with the residual shift occurring in the direction that would slightly \emph{lower}, rather than raise, the simple inverse-ruler estimate of $H_0$.

Accordingly, the present calculation does not support a sound-horizon mechanism that alleviates the Hubble tension. A precision statement about CMB-inferred parameters would additionally require a self-consistent recombination/Boltzmann calculation and a full likelihood analysis; the inverse-ruler column in Table~\ref{tab:drag_results} is included only to display the direction and approximate size of the effect.

\section{Cosmological Time Dilation in the GCT Framework}
\label{sec:CTD}

In this section, we distinguish the standard same-event proper-time dilation from the cross-epoch temporal standardization considered within the generalized cosmological time (GCT) framework~\cite{Lee:2024kxa}. We then translate the DES time-dilation measurements into the corresponding conditional parameter $b$ and outline the requirements for a future population-level likelihood.

\subsection{Proper-time propagation and GCT population standardization}

The geometrical redshift retains its standard Robertson--Walker definition,
\begin{equation}
1+z=\frac{a_{\obs}}{a_{\emit}}.
\end{equation}
For two neighboring wave crests, or any same-event-to-same-event comparison along the comoving congruence, the proper-time dilation remains
\begin{equation}
\frac{\Delta\tau_{\obs}}{\Delta\tau_{\emit}}=1+z.
\label{eq:proper_ctd}
\end{equation}
Thus a homogeneous lapse by itself does not modify the same-event proper-time propagation law.

The CTD observable considered here instead adopts a cross-epoch population-standardization hypothesis. Along the same comoving congruence let
\begin{equation}
d\tau=\mathcal N_G\,d\mathcal T_G,
\qquad
\mathcal N_G=(U^\mu\nabla_\mu\mathcal T_G)^{-1},
\end{equation}
where $\mathcal T_G$ is homogeneous on each cosmological hypersurface. For the CTD application considered here, standardized intrinsic durations are compared in this common temporal representation. With $\mathcal N_G=N(a)=a^{b/4}$, $a_{\obs}=1$, and $N_{\obs}=1$, the population-level duration ratio is
\begin{align}
\frac{\Delta\mathcal T_{G,\obs}}{\Delta\mathcal T_{G,\emit}}
&=(1+z)\frac{N_{\emit}}{N_{\obs}}
=(1+z)^{1-b/4}.
\label{eq:gct_ctd}
\end{align}
Consequently, a phenomenological fit
\begin{equation}
\Delta t_{\obs}=(1+z)^n\Delta t_{\emit}
\end{equation}
maps to
\begin{equation}
n=1-\frac b4
\label{eq:n_b_relation}
\end{equation}
under the GCT population-standardization hypothesis. Equation~\eqref{eq:n_b_relation} should be tested jointly with source evolution and selection effects; it is not a replacement for the proper-time result in Eq.~\eqref{eq:proper_ctd}.

\subsection{Current constraints from DES SNe~Ia}

The DES CTD analysis finds a combined scaling $n=1.003\pm0.005$ (stat) $\pm0.010$ (sys), consistent with the canonical $n=1$ law~\cite{DES:2024vgg}. The $i$-band fit gives $n_i=0.988\pm0.008$ (stat) and lies on the $n<1$ side. For the band-wise translation below, we conservatively assign the representative DES stretch-evolution systematic $\sigma_{n,\textrm{sys}}\simeq0.010$ to the $i$-band result; this is an adopted systematic allowance rather than a separately measured $i$-band systematic. Under the conditional mapping in Eq.~\eqref{eq:n_b_relation}, the corresponding values are
\begin{align}
b_{\rm all}&=-0.012\pm0.020\;(\textrm{stat})\pm0.040\;(\textrm{sys}),\\
b_i&=0.048\pm0.032\;(\textrm{stat})\pm0.040\;(\textrm{sys, adopted}).
\end{align}
The band-wise central value is therefore interesting as a target for future time-domain surveys, but it is not by itself robust evidence for nonzero $b$. The CTD mapping and the early-Universe acoustic ruler probe distinct sectors of the framework, so a positive value inferred from CTD does not by itself imply a reduced sound horizon or an alleviation of the Hubble tension.

\subsection{DES-calibrated forecast for future CTD tests}
\label{subsec:ctd_forecast}

A useful forecast can be constructed without assigning a toy per-object duration error or assuming that all population effects are independent. We instead calibrate the effective information directly to the published DES $i$-band CTD result, which used $N_0=1465$ SNe~Ia and obtained a statistical uncertainty $\sigma_{n,0}=0.008$ on the exponent~\cite{DES:2024vgg}. This choice is deliberately conservative for testing persistence of the $i$-band central value because it does not mix the band-specific central value with the smaller all-band uncertainty. If a future sample has a comparable redshift distribution and comparable per-object information after standardization, the statistical Fisher information scales approximately as
\begin{equation}
F_n(N)=F_{n,0}\frac{N}{N_0},
\qquad
F_{n,0}=\sigma_{n,0}^{-2},
\end{equation}
so that
\begin{equation}
\sigma_{n,\textrm{stat}}(N)
=\sigma_{n,0}\sqrt{\frac{N_0}{N}}.
\label{eq:ctd_stat_scaling}
\end{equation}
This is an empirical information-scaling forecast, not a substitute for a full population likelihood.

To illustrate the requirement, suppose that the current $i$-band central value persists,
\begin{equation}
n_\star=0.988,
\qquad
\delta_n\equiv|n_\star-1|=0.012,
\qquad
b_\star=4\delta_n=0.048.
\end{equation}
We separate a residual exponent-level population/systematic floor $\sigma_{n,\textrm{sys}}$ from the sample-size-dependent statistical term,
\begin{equation}
\sigma_n^2(N)
=\sigma_{n,\textrm{stat}}^2(N)+\sigma_{n,\textrm{sys}}^2.
\label{eq:ctd_total_error}
\end{equation}
Here $\sigma_{n,\textrm{sys}}$ should be understood as the component that does not average down as $N^{-1/2}$, for example a residual uncertainty from population evolution, band-dependent standardization, or correlated selection effects. The significance for testing $n=1$ is then
\begin{equation}
S(N)=\frac{\delta_n}{\sigma_n(N)}.
\end{equation}
Solving for the sample size required for a target significance $S$ gives
\begin{equation}
N_{\textrm{req}}(S)
=
\frac{N_0\sigma_{n,0}^2}
{\left(\delta_n/S\right)^2-\sigma_{n,\textrm{sys}}^2},
\qquad
\sigma_{n,\textrm{sys}}<\frac{\delta_n}{S}.
\label{eq:ctd_nreq}
\end{equation}
If the inequality in Eq.~\eqref{eq:ctd_nreq} is not satisfied, no finite increase in sample size can reach the requested significance. Equivalently, the asymptotic significance is $S_\infty=\delta_n/\sigma_{n,\textrm{sys}}$ for a nonzero residual floor.

\begin{table}[t]
\centering
\caption{DES-calibrated sample-size requirements for testing persistence of the illustrative $i$-band central value $n_\star=0.988$ ($b_\star=0.048$). The baseline is $N_0=1465$ and $\sigma_{n,0}=0.008$. A dash denotes that no finite sample can reach the requested significance under the stated residual exponent-level systematic floor.}
\label{tab:ctd_forecast}
\begin{tabular}{ccccc}
\toprule
$\sigma_{n,\textrm{sys}}$ & $\sigma_{b,\textrm{sys}}$ & $N_{3\sigma}$ & $N_{5\sigma}$ & $S_\infty$ \\
\midrule
0       & 0     & 5860  & 16278 & $\infty$ \\
0.002   & 0.008 & 7814  & 53273 & 6.0 \\
0.003   & 0.012 & 13395 & --    & 4.0 \\
0.004   & 0.016 & --    & --    & 3.0 \\
0.010   & 0.040 & --    & --    & 1.2 \\
\bottomrule
\end{tabular}
\end{table}

\begin{figure}[t]
\centering
\includegraphics[width=0.72\textwidth]{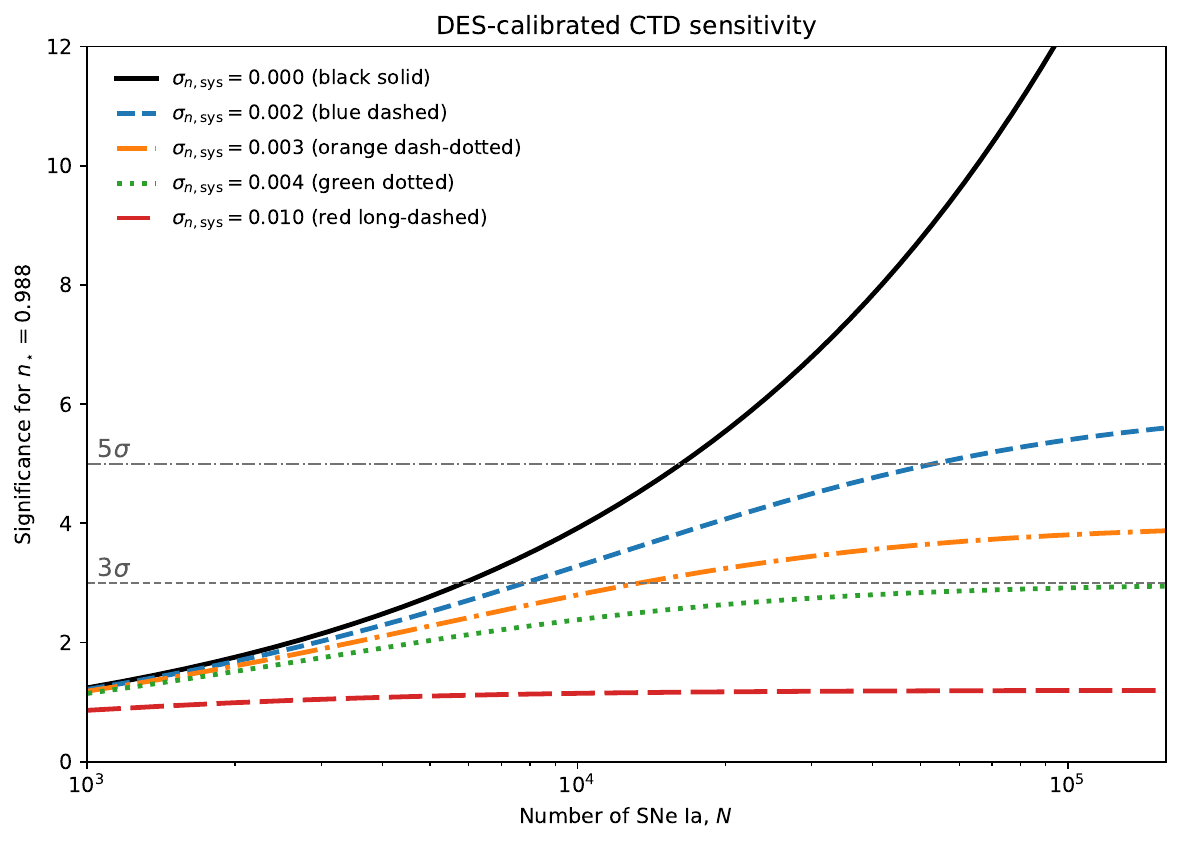}
\caption{DES-calibrated significance for testing persistence of the illustrative $i$-band central value $n_\star=0.988$ as a function of sample size. The statistical term is normalized to $N_0=1465$ and $\sigma_{n,0}=0.008$. Distinct residual exponent-level systematic floors are indicated by both color and line style: black solid ($\sigma_{n,\textrm{sys}}=0$), blue dashed ($0.002$), orange dash-dotted ($0.003$), green dotted ($0.004$), and red long-dashed ($0.010$). The horizontal guides mark the $3\sigma$ and $5\sigma$ levels. The forecast is an information-scaling sensitivity estimate and does not replace an explicit population-evolution and covariance likelihood.}
\label{fig:ctd_forecast}
\end{figure}

Figure~\ref{fig:ctd_forecast} shows the corresponding significance as the sample size grows, with each residual floor encoded by both color and line style: the black solid curve denotes the statistics-only limit $\sigma_{n,\textrm{sys}}=0$, the blue dashed curve denotes $\sigma_{n,\textrm{sys}}=0.002$, the orange dash-dotted curve denotes $\sigma_{n,\textrm{sys}}=0.003$, the green dotted curve denotes $\sigma_{n,\textrm{sys}}=0.004$, and the red long-dashed curve denotes $\sigma_{n,\textrm{sys}}=0.010$. This visualization makes the role of the residual floor immediate: each curve initially rises with increasing $N$ while the statistical term dominates, and then flattens once the asymptotic ceiling $S_\infty=\delta_n/\sigma_{n,\textrm{sys}}$ is reached. In the statistics-only limit (black solid), persistence of $n_\star=0.988$ would require approximately $5.9\times10^3$ SNe~Ia for a $3\sigma$ distinction from $n=1$ and $1.63\times10^4$ for $5\sigma$, under the same effective information per object as the present DES $i$-band analysis. The blue dashed curve still crosses both the $3\sigma$ and $5\sigma$ reference levels, whereas the orange dash-dotted curve saturates at $4\sigma$ and thus cannot reach $5\sigma$. The green dotted curve asymptotes to $3\sigma$, marking the practical boundary for a nominal $3\sigma$ test, while the red long-dashed curve remains limited to about $1.2\sigma$ even at very large $N$. Equivalently, a $3\sigma$ test requires $\sigma_{n,\textrm{sys}}<0.004$ (or $\sigma_{b,\textrm{sys}}<0.016$), while a $5\sigma$ test requires $\sigma_{n,\textrm{sys}}<0.0024$ (or $\sigma_{b,\textrm{sys}}<0.0096$).

The central lesson is different from a purely sample-size forecast. Increasing the number of transients is useful only after band-dependent source evolution, population standardization, selection effects, and correlated calibration uncertainties are controlled at the few-$10^{-3}$ level in the fitted exponent. Therefore, future Rubin/LSST analyses should fit $n$ jointly with an explicit population-evolution model and a covariance description, and only then apply Eq.~\eqref{eq:n_b_relation} under the stated GCT population-standardization hypothesis. The forecast above quantifies the precision target; it does not treat the present $i$-band central value as a detection or assume that it will persist.

\section{Conclusion}
\label{sec:conclusion}

We have investigated the relation between cosmological time dilation and the Hubble tension in the meVSL framework, with the temporal interpretation formulated through generalized cosmological time (GCT). The homogeneous temporal scaling leaves the same-event proper-time law $\Delta\tau_{\obs}/\Delta\tau_{\emit}=1+z$ unchanged, while the relation $n=1-b/4$ follows conditionally when intrinsic durations are standardized across epochs in GCT. The baryon drag epoch, by contrast, is determined by local photon--baryon momentum transfer and is governed by the Thomson drag weighting $1/R$.

Keeping the rapid $X_e(z)$ dependence in the photon-decoupling condition gives only a mild response of $z_\ast$ to $b$, locally represented by $z_\ast\simeq1090-315b$ over the range studied here. Positive $b$ also moves $z_{\drag}$ slightly downward. For a GR-like recombination history with the specified meVSL Thomson-cross-section scaling, the resulting sound horizon remains close to its standard value and increases slightly, from about $147.15$ Mpc at $b=0$ to $147.84$ Mpc at $b=0.03$ in the simplified calculation.

The early-Universe calculation therefore does not provide the several-percent reduction of $r_{\drag}$ required for a sound-horizon resolution of the Hubble tension. A simple inverse-ruler scaling moves $H_0$ slightly downward rather than upward, although a precision cosmological constraint requires a full Boltzmann and likelihood analysis. CTD remains an independent empirical test of GCT cross-epoch temporal standardization and need not track the behavior of the early-Universe acoustic ruler.

A precision treatment should implement any physical $b$ dependence of atomic microphysics self-consistently in a HyRec/CLASS-type calculation and analyze transient CTD with an explicit source-evolution and covariance model. These observables test different aspects of the framework and can be confronted with data separately before a combined interpretation is attempted.

\appendix

\section{Recombination history}
\label{app:Xe}

The early-Universe calculation takes the hydrogen recombination history to be unchanged from GR, while the effective Thomson cross section in the homogeneous cosmological opacity representation retains its stated $b$ dependence. To estimate the displacement of the relevant epochs we use a standard effective three-level Peebles hydrogen recombination history~\cite{Peebles:1968ja} with the fiducial physical densities $\Omega_mh^2=0.1423$ and $\Omega_bh^2=0.02237$, $h=0.6736$, $T_{\rm CMB}=2.7255$ K, $N_{\rm eff}=3.046$, and $Y_p=0.245$. Representative values are listed in Table~\ref{tab:Xe}.
\begin{table}[htbp]
\centering
\begin{tabular}{cc}
\toprule
$z$ & $X_e^{\GR}(z)$\\
\midrule
1200 & 0.32613\\
1150 & 0.22546\\
1100 & 0.14543\\
1090 & 0.13206\\
1060 & 0.09713\\
1050 & 0.08716\\
1000 & 0.04855\\
900  & 0.01279\\
\bottomrule
\end{tabular}
\caption{Representative free-electron fractions used to characterize the recombination transition. These values are adequate for estimating the epoch shifts considered here but are not a substitute for a precision HyRec/CLASS recombination history.}
\label{tab:Xe}
\end{table}
The baseline optical-depth normalization is anchored to $z_\ast=1090$ and $z_{\drag}=1059.94$ at $b=0$. This procedure is intended to isolate the response to the stated meVSL background-opacity scaling. If one instead modifies local atomic binding energies and all associated microscopic rates, a closed $b$-dependent recombination network must be specified and solved separately.

\section{Local response of the photon-decoupling redshift}
\label{app:zstar}

Let $y=1+z_\ast$ and $y_{\rm eq}=1+z_{\rm eq}$. Equation~\eqref{eq:zstar_condition} can be written as
\begin{equation}
F(y,b)=\ln C+\ln X_e(y)+\frac{3+b}{2}\ln y
-\frac12\ln\!\left(1+\frac{y}{y_{\rm eq}}\right)=0,
\end{equation}
where $C$ contains the redshift-independent factors. Implicit differentiation gives
\begin{equation}
\frac{dy}{db}
=-\frac{\tfrac12\ln y}
{\dfrac{d\ln X_e}{dy}+\dfrac{3+b}{2y}-\dfrac{1}{2(y_{\rm eq}+y)}}.
\label{eq:dydb}
\end{equation}
At $b=0$ and $z_\ast\simeq1090$, the numerical recombination history gives $d\ln X_e/dz\simeq9.8\times10^{-3}$, which is substantially larger than the remaining expansion terms in the denominator. The resulting local response is $dz_\ast/db\simeq-3.2\times10^2$, consistent with the numerical fit in Eq.~\eqref{tzb}. The $X_e$ derivative is therefore the leading contribution that controls the local response of the decoupling redshift.

\section{Baryon drag optical depth and numerical interpolation}
\label{app:zdrag}

For the conventional definition $R=3\rho_b/(4\rho_\gamma)$, the baryon Euler equation contains the standard Thomson momentum-transfer term~\cite{Ma:1995ey,Hu:1995en}. In our notation this term is proportional to $\dot\kappa(\theta_\gamma-\theta_b)/R$. The corresponding positive drag optical depth in redshift is
\begin{equation}
\tau_{\drag}(z;b)=
\int_0^z dz'\,\frac{1}{R(z')}
\frac{X_e(z')n_{b0}\sigma_{\rm T}c_0}{H(z')}
(1+z')^{2+b/2},
\label{eq:taudrag_appendix}
\end{equation}
with $\tau_{\drag}(z_{\drag};b)=1$. The numerical nodes quoted in Eq.~\eqref{eq:zdrag_nodes} are generated from this definition. A quadratic fit to the extended node set $0\le b\le0.08$ gives Eq.~\eqref{eq:zdrag_fit_exp_final}; its root-mean-square interpolation residual is below $10^{-3}$ in redshift for the numerical node table.

The corresponding sound horizon is evaluated from
\begin{equation}
r_{\drag}(b)=\frac{c_0}{\sqrt3}
\int_{z_{\drag}(b)}^\infty\frac{dz}{H(z)}
\left[1+\frac{1+b/6}{1+b/8}R(z)\right]^{-1/2},
\label{eq:rdrag_appendix}
\end{equation}
which is the same acoustic integral used in the main text after the $\tilde c/\tilde H$ cancellation.

%%%%%%%%%%%%%%%%%%%%%%%%%%%%%%%%%%%%%%%%%%%
%%%%%%%%%%%%%%%%%%%%%%%%%%%%%%%%%%%%%%%%%%%

%\bibliography{refs}

\end{document}